\documentclass[
twocolumn,
superscriptaddress,
amsmath,amssymb,
prl,
]{revtex4-2}
\usepackage{graphicx}
\usepackage{hyperref}
\usepackage{dcolumn}
\usepackage{bm}
\usepackage{epstopdf}
\usepackage{romannum}
\usepackage{csquotes}
\usepackage[dvipsnames]{xcolor}
\usepackage{hyperref}

\begin{document}

\preprint{La0.8Sr0.2NiO2}

\title{Large upper critical fields and dimensionality crossover of superconductivity in infinite-layer nickelate La$_{0.8}$Sr$_{0.2}$NiO$_{2}$}

\author{Wei Wei}
\affiliation{Department of Physics, Southeast University, Nanjing 211189, China}

\author{Wenjie Sun}
\affiliation{National Laboratory of Solid State Microstructures, Jiangsu Key Laboratory of Artificial Functional Materials, College of Engineering and Applied Sciences, Nanjing University, Nanjing 210093, P. R. China.}
\affiliation{Collaborative Innovation Center of Advanced Microstructures, Nanjing University, Nanjing 210093, P. R. China.}

\author{Yue Sun}

\email{Corresponding author: sunyue@seu.edu.cn}
\affiliation{Department of Physics, Southeast University, Nanjing 211189, China}

\author{Yongqiang Pan}
\affiliation{Key Laboratory of Materials Physics, Institute of Solid State Physics,
	Chinese Academy of Sciences, Hefei, 230031, China}

\author{Gangjian Jin}
\affiliation{School of Electrical and Electronic Engineering, Huazhong University of Science and Technology, Wuhan 430074, China}
\author{Feng Yang}
\affiliation{School of Physics, Huazhong University of Science and Technology, Wuhan 430074, China}

\author{Yueying Li}
\affiliation{National Laboratory of Solid State Microstructures, Jiangsu Key Laboratory of Artificial Functional Materials, College of Engineering and Applied Sciences, Nanjing University, Nanjing 210093, P. R. China.}
\affiliation{Collaborative Innovation Center of Advanced Microstructures, Nanjing University, Nanjing 210093, P. R. China.}

\author{Zengwei Zhu}
\affiliation{Wuhan National High Magnetic Field Center and School of Physics, Huazhong University of Science and Technology, Wuhan 430074, China}

\author{Yuefeng Nie}
\email{Corresponding author:ynie@nju.edu.cn}
\affiliation{National Laboratory of Solid State Microstructures, Jiangsu Key Laboratory of Artificial Functional Materials, College of Engineering and Applied Sciences, Nanjing University, Nanjing 210093, P. R. China.}
\affiliation{Collaborative Innovation Center of Advanced Microstructures, Nanjing University, Nanjing 210093, P. R. China.}
\author{Zhixiang Shi}
\email{Corresponding author:zxshi@seu.edu.cn}
\affiliation{Department of Physics, Southeast University, Nanjing 211189, China}

\begin{abstract}
 
\textbf{}
The recently emerging superconductivity in infinite-layer nickelates, with isostructure and isoelectron of cuprates, provides a new platform to explore the pairing mechanism of high-temperature superconductors. In this work, we studied the upper critical field ($H_{\rm{c2}}$) of a high-quality La$_{0.8}$Sr$_{0.2}$NiO$_{2}$ thin film with superconducting transition temperature, $T_{\rm{c}}$ = 18.8 K, using high magnetic field up to 56 T. A very large $H_{\rm{c2}}$,  $\sim$ 40 T for $H$ $\Arrowvert$ $c$ and $\sim$ 52 T for $H$ $\Arrowvert$ $ab$, was confirmed, which suggests that infinite-layer nickelates also have great application potential. The anisotropy of $H_{\rm{c2}}$ monotonically decreases from $\sim$ 10 near $T_{\rm{c}}$ to $\sim$ 1.5 at 2 K. Angle dependence of $H_{\rm{c2}}$ confirms the crossover of superconductivity from two-dimensional (2D) to three-dimensional (3D) as the temperature decreases. We discussed that the interstitial orbital effect causes the weakening of anisotropy. The observed abnormal upturning of $H_{\rm{c2}}$ at low temperatures is found to be a universal behavior independent of film quality and rare earth elements. Therefore, it should not be the Fulde-Ferrell-Larkin-Ovchinnikov (FFLO) state due to the fact that it is in the dirty limit and insensitive to disorder. 
\end{abstract}

 
\maketitle
\textbf{}
Research of the high temperature superconducting (SC) mechanism and possible contributions involved in the pairing process is the foundation for realizing the dream of high-temperature superconductivity (HTSC). Understanding electronic properties of materials similar to the cuprates can provide insight into the mechanism of superconductivity in the cuprates and other high-$T_{\rm{c}}$ superconductors. The infinite-layer nickel-based 112 thin films consist of square plane-coordinated NiO$_{2}$ planes and Ln/Sr/Ca atomic layers, which is the isostructure of cuprates.\cite{1} For the LaNiO$_{2}$ and NdNiO$_{2}$, X-ray absorption spectroscopy (XAS) and resonant inelastic X-ray scattering (RIXS) confirmed the nominal 3 $d^{9}$ electronic configuration, equivalent to Cu$^{2+}$ in the cuprates.\cite{2} Such isostructural and isoelectronic characteristics show great promise in deconstructing the various cooperative mechanisms responsible for HTSC. However, different from cuprates, the hybridization of Ni 3$d$ and O 2$p$ orbitals decreases and the coupling of Ni 3$d$ and La/Nd 5$d$ states increases.\cite{2} 
In this configuration, the Ni state can have a charge transfer to the rare earth cation, thus leaving holes in Ni orbitals and electrons in $R$ 5$d$ ($ R $ = La, Pr, Nd) orbitals, which is known as the self-doping effect.\cite{3,63} The change in sign of the low-temperature Hall coefficient as a function of Sr doping indicates the presence of multiple bands at the Fermi level.\cite{4,5} From the theoretical point of view, electronic structure calculations for the nickelates indicate the presence of a large hole pocket with 3$d_{x^{2}–y^{2}}$ character, and electron pockets arising from Nd 5$d$ and Ni 3$d$ hybridization.\cite{1,8,9} As is well known, the parent phase of cuprates is depicted as a charge transfer insulator with an antiferromagnetic (AFM) order.\cite{10,63} In sharp contrast, a definitive resolution of long range AFM in infinite-layer nickelates is still lacking, even though hints of short-range magnetic order or spin glass behavior have been picked up by various measurements.\cite{11,12,13,14} Recent STM experiments uncovered a mixed $s$- and $d$-wave SC gap feature on the rough surface of Nd$_{1-\textit{x}}$Sr$_{\textit{x}}$NiO$_{2}$ thin films.\cite{15} At present, the research on nickelates superconductors is still in the exploratory stage. The similarities and differences between the SC mechanisms of nickelates and cuprates, and the reasons for the lack of superconductivity in bulk nickelates are still unclear.

\begin{figure*}\center
	\includegraphics[width=0.7\linewidth]{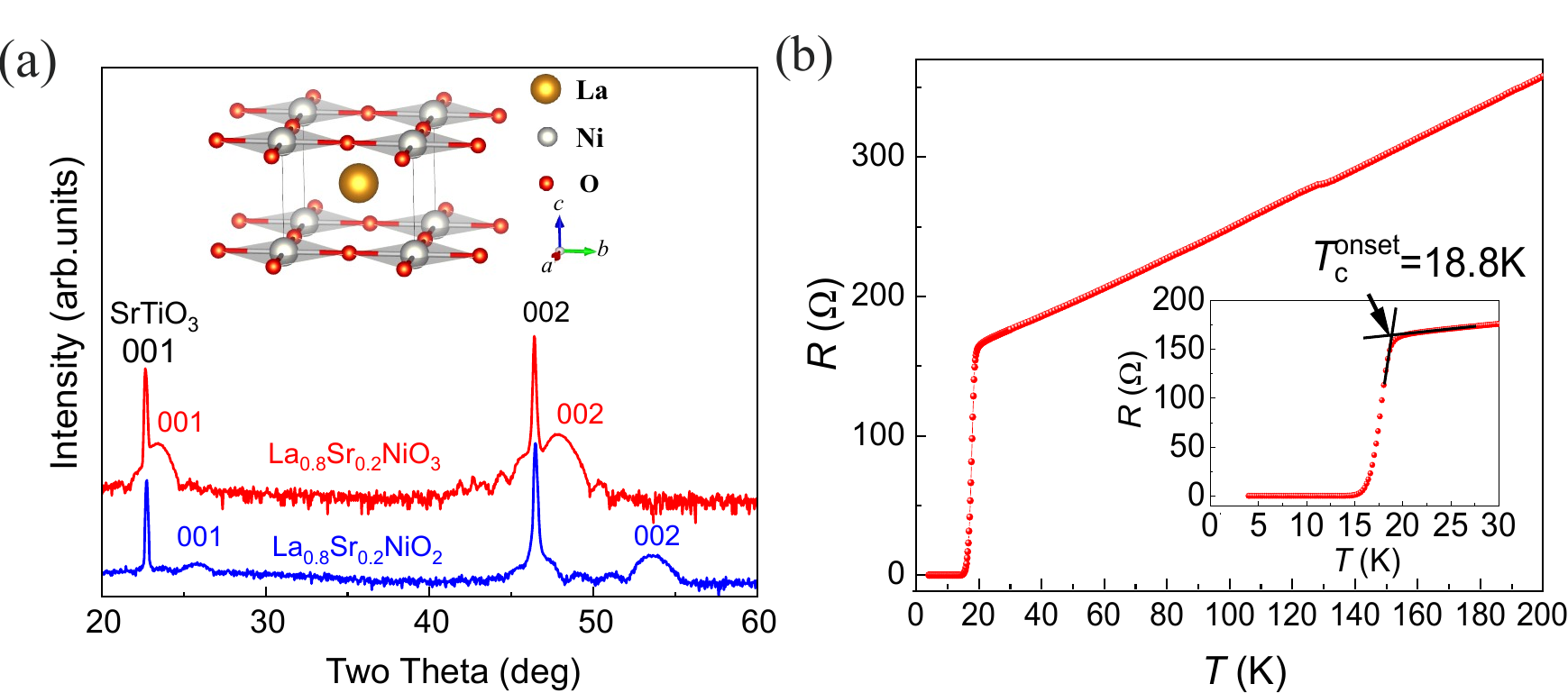}
	\caption{\textbf{(a)}  XRD patterns of the perovskite La$_{0.8}$Sr$_{0.2}$NiO$_{3}$ (red) and infinite-layer thin film La$_{0.8}$Sr$_{0.2}$NiO$_{2}$ (blue). Inset of a) shows the crystal structure of La$_{0.8}$Sr$_{0.2}$NiO$_{2}$. b) Temperature dependence of the in-plane resistance $R$($T$) of thin film. Inset: Zoom-in view of $R$($T$) near the superconducting transition, showing $T_{\rm{c}}^{\rm{onset}}$ = 18.8 K.}\label{}
\end{figure*}

The upper critical field ($H_{\rm{c2}}$) is a fundamental parameter for getting information such as pair-breaking mechanism, coherence length $\xi$, and pairing symmetry, which plays an important role in the understanding of unconventional superconducting mechanism.\cite{16,17,18} It has been found that the $H_{\rm{c2}}$ of Nd$_{0.775}$Sr$_{0.225}$NiO$_{2}$ with SC transition temperature ($T_{\rm{c}}$) of 8.5 K is surprisingly isotropic at low temperatures despite the layered crystal structure.\cite{19} In addition, $H_{\rm{c2}}$ at 0 K for $H_{\rm{c2}}^{ab}$ ($ H $ $ \Arrowvert $ $ ab $) and $H_{\rm{c2}}^{c}$ ($ H $ $ \Arrowvert $ $ c $) estimated by simply linear extrapolating $H_{\rm{c2}}$($T$) to 0 K are relatively small $\sim$ 14 T and 11 T. At the same time, a prominent low-temperature upturn in $H_{\rm{c2}}$($T$) has been observed.\cite{19} Superconductivity of La-, Pr- and Nd-nickelates has also been investigated.\cite{22,23,24} The unique polar and azimuthal angle-dependent magnetoresistance observed in Nd-nickelate can be understood as a magnetic contribution from Nd$^{3+}$ 4$f$ moments, regardless of the specific doping.\cite{25} It is also striking that $H_{\rm{c2}}$ of the La- and Pr-nickelate thin film substantially exceeds the conventional Pauli limit.\cite{20,21,25} Therefore, the study of La- and Pr-nickelates without the influence from 4$f$ moments can reveal the intrinsic properties of NiO$_{2}$ plane.

Until now, the reported values of $H{\rm_{c2}}$ for infinite layer nickelates are still relatively small, much lower than those of cuprates and Iron-based superconductors (IBSs).\cite{16,29,30} Is the small $H_{\rm{c2}}$ the intrinsic characteristic of infinite layer nickelates? Is the upturning behavior of $H{\rm_{c2}}$ at low temperature related to rare earth elements or disorder? In this letter, we studied the temperature and polar angle dependence of $H{\rm_{c2}}$ in a high-quality La$_{0.8}$Sr$_{0.2}$NiO$_{2}$ thin film with $T_{\rm{c}}$ $\sim$ 18.8 K, by measuring resistance under a pulsed magnetic field up to 56 T. The $H_{\rm{c2}}$ at 0 K for $H_{\rm{c2}}^{ab}$ and $H_{\rm{c2}}^{c}$ are estimated to be $\sim$ 52 T and $\sim$ 40 T, which are the largest among infinite-layer nickelate superconductors reported so far. In addition, the angle-dependent measurements of $H_{\rm{c2}}$ reveal the crossover of superconductivity from 2D to 3D. 


Thin film of the precursor La$_{0.8}$Sr$_{0.2}$NiO$_{3}$ with thickness of $\sim$ 6.8 nm was grown by molecular beam epitaxy (MBE) on the TiO$_{2}$ - terminated SrTiO$_{3}$ (001) substrate, with substrate temperature of 600 $^\circ$C and an oxidant background pressure of $\sim$ 1 × 10$^{-5}$ Torr using the distilled ozone during the reaction. For stoichiometry and growth parameters, we used the optimal conditions previously reported. To obtain the infinite-layer structure, the mixtures of precursor and CaH$_{2}$ powder ( $\sim$ 0.1 g) were sintered at 310 $^\circ$C for 4 h, with warming (cooling) rate of 10 – 15$^\circ$C/min. 
The film structure was characterized by X-ray diffraction (XRD) with Cu-K$\alpha$ radiation on a Bruker D8 Discover diffractometer. Electrical resistance was measured up to 9 T on a Quantum Design Physical Property Measurement System (PPMS - 9 T) and in pulsed field up to 56 T at Wuhan National High Magnetic Field Center using a standard four-probe technique.
\textbf{}

\begin{figure*}\center
	\includegraphics[width=1\linewidth]{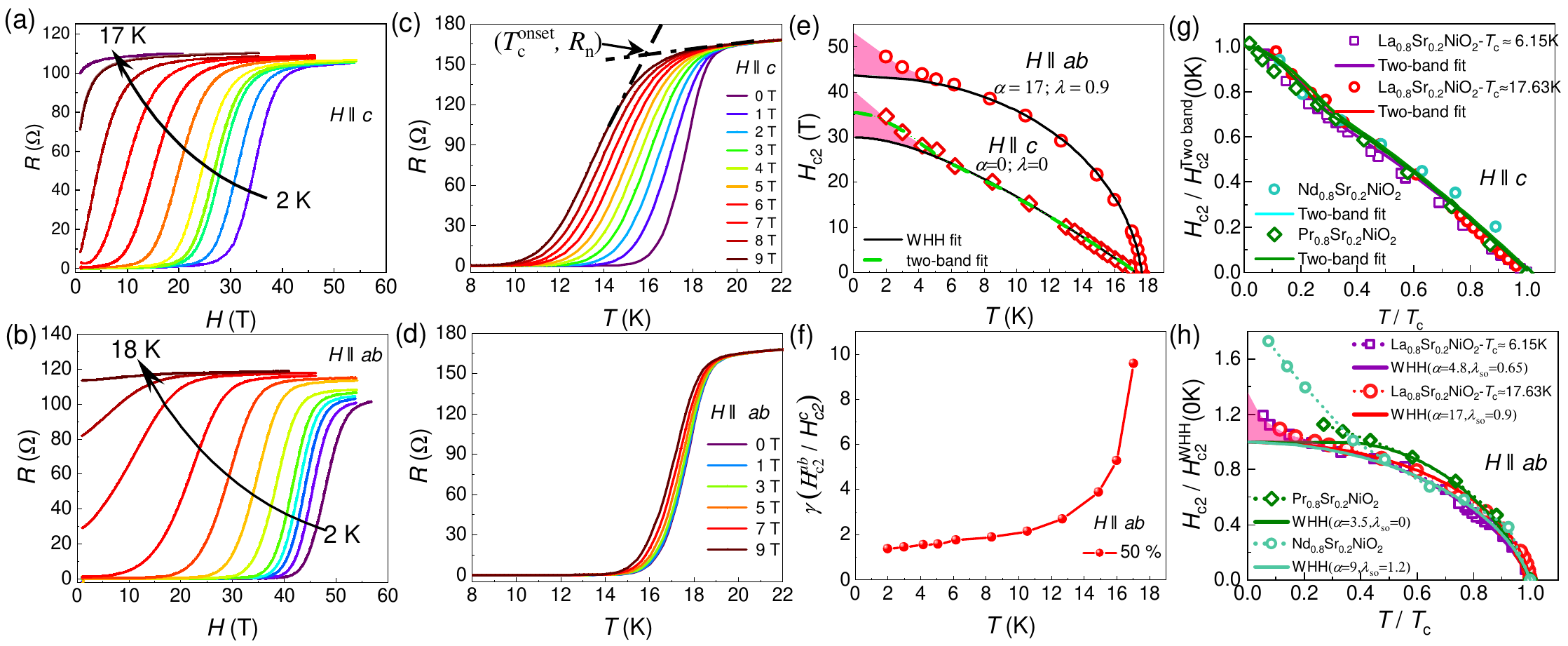}
	\caption{The magnetic field dependence of resistance for La$_{0.8}$Sr$_{0.2}$NiO$_{2}$ with \textbf{(a)} $H$ $\Arrowvert$ $c$ at 2, 3, 4.2, 5, 6, 8, 10, 12, 14, 16, and 17 K, \textbf{(b)} $H$ $\Arrowvert$ $ab$ at 2, 3, 4.2, 5, 6, 8, 10, 12, 14, 16, 17, and 18 K.
	Temperatures dependence of resistance for \textbf{(c)} $H$ $\Arrowvert$ $c$ and \textbf{(d)} $H$ $\Arrowvert$ $ab$ under magnetic fields from 0 to 9 T. \textbf{(e)} $H$$_{\rm{c2}}$ represents the upper critical field for $H$ $\Arrowvert$ $c$ and $H$ $\Arrowvert$ $ab$ determined by taking a criterion of 50$\%$ $R_{\rm{n}}$; Black and green lines are WHH and two-band fitting curves, respectively. \textbf{(f)} Anisotropy of the upper critical field $\gamma$ = $H_{\rm{c2}}^{ab}$ / $H_{\rm{c2}}^{c}$. \textbf{(g-h)} The $H_{\rm{c2}}$ of R$_{0.8}$Sr$_{0.2}$NiO$_{2}$ (R = La, Pr or Nd) under \textbf{(g)} $H$ $\Arrowvert$ $ab$, \textbf{(h)} $H$ $\Arrowvert$ $c$ are derived from the report of Harold Y. Hwang $ et$ $al $.. The obtained data was normalized by $H_{\rm{c2}}^{\rm{WHH}}$ (0 K) and $H_{\rm{c2}}^{\rm{Two-band}}$(0 K) for comparison. Solid lines represent WHH and two-band fit. The pink area represents tne region where anomalous upturns of $H_{\rm{c2}}$ are observed.}\label{}
\end{figure*}
Figure 1a shows the XRD patterns of the perovskite La$_{0.8}$Sr$_{0.2}$NiO$_{3}$ (red) and infinite-layer thin film La$_{0.8}$Sr$_{0.2}$NiO$_{2}$ (blue), with only (00$l$) peaks observed. The illustration shows the structure of La$_{0.8}$Sr$_{0.2}$NiO$_{2}$. The shift of (00$l$) peaks indicate the transition of perovskite structure to infinite layer structure. Furthermore, we have studied atomic arrangement before using high-angle annular dark-field scanning transmission electron microscopy (HAADF-STEM) and energy-dispersive X-ray spectroscopy (EDS).\cite{21} The Ruddlesden-Popper stacking faults are not observed. Highly ordered arrangements of La, Sr and Ni atoms and sharp interfaces are also observed without obvious defects such as cation intermixing and stacking. The high quality of our thin film is also reflected in the significant enhancement of $T_{\rm{c}}$. The value of $T_{\rm{c}}$ is $\sim$ 18.8 K, as shown in Figure 1d, which is more than twice larger than that of $\sim$ 8 K in previous reports.\cite{25,46,22}

Temperature dependence of in-plane resistance with $H$ $\Arrowvert$ $c$ and $H$ $\Arrowvert$ $ab$ under fields from 0 to 9 T are shown in Figure 2c-d. The resistive transition shifts towards lower $T$ under higher fields. Small broadening is recognized with increasing $H$, particularly for $H$ $\Arrowvert$ $ab$. Figure 2a-b show the in-plane resistance of La$_{0.8}$Sr$_{0.2}$NiO$_{2}$ thin film as a function of the magnetic fields. We obtain the $H^{ab}_{\rm{c2}}$ (circle symbol) and $H^{c}_{\rm{c2}}$ (diamond symbol) in Figure 2e by the criterion of 50$\%$ of the normal-state resistance values $R$$_{n}$, where the determination of $R$$_{n}$ is shown in Figure 2c.   
Obviously, we observe a linear temperature dependence of $H_{\rm{c2}}^{c}$ and a square root temperature dependence of $H_{\rm{c2}}^{ab}$ (see Figure 2e). Similar nearly linear $H^{c}_{\rm{c2}}$-$T$ has been observed in most IBSs, suggesting that the spin paramagnetic effect is negligible when $H$ $\Arrowvert$ $c$ (or much smaller than $H$ $ \Arrowvert $ $ab$) \cite{31,32,33,26}. In addition, the $H_{\rm{c2}}$ at 0 K for $H^{ab}_{\rm{c2}}$ and $H^{c}_{\rm{c2}}$ estimated by simply linear extrapolating $H_{\rm{c2}}$($T$) to 0 K are $ \sim$ 52 T and 40 T, which are the largest among the reported values in infinite-layer nickelate superconductors. Such value of $H_{\rm{c2}}$ is close to that of IBSs “122” sample,\cite{26,27} which indicates the application potential for the Ni-based superconductors.
\begin{figure*}\center
	\includegraphics[width=1\linewidth]{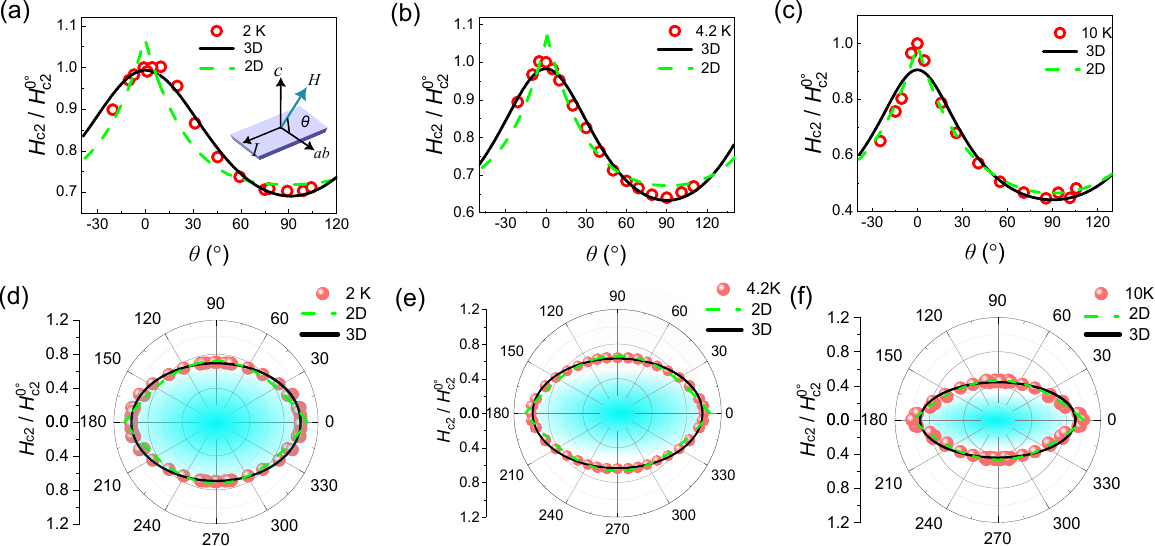}
	\caption{\textbf{(a-c)} Polar angle dependence of $H_{\rm{c2}}$($\theta$) at \textbf{(a)} 2 K, \textbf{(b)} 4.2 K and \textbf{(c)} 10 K, respectively. The black solid lines and the green dashed lines are the fitting by 3D GL model and 2D Tinkham model. Inset of \textbf{(a)} shows schematic structure for measurement. Polar plots of the angle-dependent upper critical field at \textbf{(d)} 2 K, \textbf{(e)} 4.2 K and \textbf{(f)} 10 K,respectively.}\label{}
\end{figure*}

In weakly coupled BCS superconductors, $H_{\rm{p}}$ (= 1.86 $ \times $ $T_{\rm{c}}$) is the Pauli-limited field, which characterises the binding energy of Cooper pairs.\cite{34} We observed the violations of Pauli-limit ($H_{\rm{p}}$ = 32.38 T) in all directions in our work. Ariando $et$ $al.$ observed similar behavior of violating Pauli-limit in the La-nickelates with low $T_{\rm{c}}$ \cite{20}. For a spin-singlet superconductor, $H_{\rm{c2}}$ is usually smaller than $H_{\rm{p}}$, except for some special cases such as the finite-momentum pairing \cite{35}, strong spin-orbit-coupling,\cite{36,37} large pseudogap or strong-coupling exist. Werthamer, Helfand, and Hohenberg (WHH) theory is the most commonly used method to study the behavior of $H_{\rm{c2}}$, taking into account both the spin paramagnetic effect, and the spin-orbit scattering effect.\cite{38} The magnitude of spin paramagnetic effect and spin-orbit scattering are expressed by $\alpha$ and $\lambda$$_{so}$, respectively. $H^{c}_{\rm{c2}}$, exhibiting a quasi-linear temperature dependence, can be described better by the two-band model, except for the anomalous upturn at low temperatures. $H^{ab}_{\rm{c2}}$ can be fitted well by the WHH mode with $\alpha$ = 17 and $ \lambda $$_{so}$ = 0.9 at $T$ $>$ 5 K. It is interesting to note that when $T$ $<$ 5 K, anomalous upturns of $H^{c}_{\rm{c2}}$ and $H^{ab}_{\rm{c2}}$ beyond the upper limit expected from WHH theory are observed. For comparison, $H_{\rm{c2}}^{ab}$ normalized by $H_{\rm{c2}}^{\rm{WHH}}$ (0K) and $H_{\rm{c2}}^{c}$ normalized by $H_{\rm{c2}}^{\rm{Two-band}}$ (0K) are shown in Figure 2g-h, respectively. Similar upturn behaviors have also been observed in La$_{0.8}$Sr$_{0.2}$NiO$_{2}$ with $T_{\rm{c}}$ = 6.15 K (Purple rectangle), Nd$_{0.8}$Sr$_{0.2}$NiO$_{2}$ (Cyan Circle) and Pr$_{0.8}$Sr$_{0.2}$NiO$_{2}$ (Green diamond).\cite{25}

The anomalous upturns observed in heavy-fermion,\cite{35,39,41} organic superconductors,\cite{42,43} and some IBSs\cite{44} have been discussed to originate from the FFLO state. For conventional $s$-wave superconductors, the FFLO state is readily destroyed by impurities.\cite{35} In La$_{0.8}$Sr$_{0.2}$NiO$_{2}$ thin films with different $T_{\rm{c}}$ (i.e. different degrees of disorder), the degree of upturn of $H_{\rm{c2}}$/$H_{\rm{c2}}^{\rm{WHH}}$ (0K) is almost the same in the scaled plot (see Figure 2g). Therefore, the high-field SC phase is not related to the degree of disorder. On the other hand, our previous report has estimated that the coherence length $\xi$$_{0}$ is $\sim$ 127.6 $\rm\AA$, much larger than the carrier mean free path $ \lambda $$_{mfp}$ of 8.2 $\rm\AA$, indicating that the film is in the dirty limit.\cite{21} These evidences suggest that $H_{\rm{c2}}$ upturn in nickelate superconductors are not caused by FFLO state. 
Furthermore, the disorder-independent upturn can also exclude the possibility of Anderson localization effect, but a clean-limit sample is needed for further investigation. 
Another possible origin of the high-field SC phase is the spin-density wave (SDW). In the heavy-fermion superconductor CeCoIn$_{5}$, an additional SDW was observed to coexist with superconductivity.\cite{49,50} Hybridization of La 5$d$ and Ni 3$d$ orbital leads to the enhancement of interlayer electron coupling with the decrease of temperature, which may induce the fluctuation of SDW. However, more evidence is needed for the coexistence of SDW and superconductivity in nickelate. The multiple bands effect may also be the origin of low-temperature upwarping, which have been fully discussed in Nd$ _{0.775} $Sr$ _{0.225} $NiO$ _{2} $ thin film.\cite{19} Actually, $H^{c}_{\rm{c2}}$ ($T$) curve could be fitted by the two-band model (see Figure 2e). Until now, the microscopic origin of the high-field phase is still open to debate. 

As shown in Figure 2f, the anisotropy $\gamma$ (= $H_{\rm{c2}}^{ab}$ / $H_{\rm{c2}}^{c}$) monotonically decreases with decreasing temperatures from $\sim$ 10 near $T_{c}$ to $\sim$ 1.5 at low temperatures. The value of $ \gamma $ in our high-quality thin film is slightly larger than the previous reports with lower $T_{\rm{c}}$ (7 for La$_{0.8}$Sr$_{0.2}$NiO$_{2}$ and 6 for La$_{0.8}$Ca$_{0.2}$NiO$_{2}$ near $T_{\rm{c}}$).\cite{25} It is worth noting that the large anisotropy observed in La$_{0.8}$Sr$_{0.2}$NiO$_{2}$ is different from that of Nd$_{0.775}$Sr$_{0.225}$NiO$_{2}$. For Nd$_{0.775}$Sr$_{0.225}$NiO$_{2}$, isotropic $H_{\rm{c2}}$ has been obtained despite the layered crystal structure, quasi-2D band structure ($d$$_{x^{2}-y^{2}}$) and the thin film geometry. Regarding the Nd$^{3+}$ 4$f$ moment’s easy-plane magneto-anisotropy, the Nd$^{3+}$ magnetic susceptibility reaches a peak in plane.\cite{25} This leads to a maximum amplification of local in-plane magnetic field and consequently a stronger suppression of superconductivity. Therefore, the isotropic upper critical field is observed in Nd$_{0.775}$Sr$_{0.225}$NiO$_{2}$.\cite{19} 
First principle calculations in LaNiO$_{2}$ found two small electron-like pockets around the $\Gamma$ and $A$ points.\cite{62,63} These electronic pockets are contributed mainly from La-derived orbitals.\cite{63} At about 20$\%$-hole doping, the electron-like pocket around the $\Gamma$ point nearly disappears, leaving only a diminished three-dimensional (3D) electron pocket around the $A$ point.\cite{61} Hence, the anisotropy observed in high-quality La$_{0.8}$Sr$_{0.2}$NiO$_{2}$ thin films may be an essential feature of infinite-layer nickelate films free from the influence of Nd$^{3+}$ magnetism. With the decrease of temperatures, orbital hybridization leads to the enhancement of electron coupling strength and the decrease of anisotropy.
 
To further study the anisotropy, angle dependence of $H_{\rm{c2}}$ at 2 K, 4.2 K, and 10 K was measured. Fig. 3 shows the $H_{\rm{c2}}$$(\theta)$ at different temperatures extracted from Figure S1. The polar angle $\theta$ is defined as the angle between the magnetic field and the $ab$ plane. For 3D bulk superconductors, the angle dependence of $H_{\rm{c2}}$ can be interpreted by the Ginzburg–Landau (GL) model with formula\cite{52,54}
\begin{eqnarray*}
	\\H_{c2}(\theta)=H_{c2}^{\Arrowvert} / \sqrt{\cos^{2}\theta+(H_{c2}^{\Arrowvert} / H_{c2}^{\perp})^{2}\sin^{2}\theta},
\end{eqnarray*}
where $H_{\rm{c2}}^{\perp}$ ($H_{\rm{c2}}^{\Arrowvert}$) is the upper critical fields with the field perpendicular (parallel) to the crystal surface. For sufficiently thin films with thickness $d$ satisfying $d$ $\ll$ $\xi$$_{c}$ ($\xi$$_{c}$ is the $c$-axis coherence length), the Tinkham formula is satisfied, which is expressed as\cite{53,54} 
\begin{eqnarray*}
	\\\left|\frac{H_{c2}(\theta)sin\theta}{H_{c2}^{\perp}}\right|+\left(\frac{H_{c2}(\theta)cos\theta}{H_{c2}^{\Arrowvert}}\right)^2=1.
\end{eqnarray*}
Obviously, the 2D Tinkham model fits $H_{\rm{c2}}$($\theta$) at 10 K better than the 3D GL model. However, at 4.2 K, the angle dependence of $H_{\rm{c2}}$ deviates from the prediction of Tinkham theory. As the temperature decreased to 2 K, $ H{\rm_{c2}} $($ \theta $) conforms to the GL model, manifesting 3D superconducting characteristic. These results suggest the dimensional crossover from 2D to 3D as the temperature decreases. 

It is well known that the rare-earth 5$d$-Ni 3$d$ orbital hybridization will form an interstitial-$s$ orbital, resulting in the existence of extended $s$-wave gaps.\cite{2,56,57} The coupling strength between the NiO$_{2}$ planes and rare earth spacer planes is enhanced with the decrease of temperature through the establishment of interstitial-$s$ orbital, leading to a 2D-to-3D crossover.\cite{58,59} The multiple bands effect with 2D wave function in high-temperature region and 3D wave function in low-temperature region can also explain the upwarping of $H_{\rm{c2}}$ at low temperatures.

For a superconductor to carry the nondissipative supercurrent, the irreversibility field $H_{\rm{irr}}$, is very crucial. As shown in Figure S2, we construct the vortex phase diagram. This phase line actually separates the phase diagram into zero and finite resistive dissipation. The relatively large $H_{\rm{irr}}$ of our high-quality thin film confirms the ability to carry lossless current under strong magnetic field, which indicates the great application prospect.

In summary, we reported a high-field study of high-quality La$_{0.8}$Sr$_{0.2}$NiO$_{2}$ thin film with $T_{\rm{c}}^{\rm{onset}}$ $\sim$ 18.8 K. It exhibits large $H_{\rm{c2}}$ ($H^{c}_{\rm{c2}}$(0 K) $\sim$ 40 T, $H^{ab}_{\rm{c2}}$(0 K) $\sim$ 52 T) and $H_{\rm{irr}}$. This combination of high $T_{\rm{c}}$, high $H_{\rm{c2}}$ and high $H_{\rm{irr}}$ is highly desirable for applications. The anisotropy $\gamma$ of $H_{\rm{c2}}$ decreases monotonically from $\sim$ 10 near $T_{\rm{c}}$ to $\sim$ 1.5 at low temperatures. The evolution of anisotropy is confirmed by the angle dependence measurements of $ H{\rm_{c2}} $, which indicates a crossover from 2D to 3D superconducting state. Hybridization of La 5$d$ and Ni 3$d$ orbital leads to the enhancement of interlayer electron coupling with the decrease of temperature, which is thought to be the possible origin of the abnormal upturning of $ H{\rm_{c2}} $ at low temperatures and 2D-to-3D crossover. These results can help us to compare the nickelates with the cuprates and to understand the high-$T_{\rm{c}}$ superconducting mechanism. 

This work was partly supported by the National Key R$\& $D Program of China (Grant No. 2018YFA0704300), the Strategic Priority Research Program (B) of the Chinese Academy of Sciences (Grant No. XDB25000000), the National Natural Science Foundation of China (Grants No. U1932217).

W.W., W.S. and Y.S. contributed equally to this paper.

\acknowledgements

\bibliographystyle{unsrt}%
\bibliography{ref}

\pagebreak

\newpage
\onecolumngrid
\begin{center}
	\textbf{\huge Supplemental information}
\end{center}
\vspace{1cm}
\twocolumngrid
\setcounter{equation}{0}
\setcounter{figure}{0}
\setcounter{table}{0}

\makeatletter
\renewcommand{\theequation}{S\arabic{equation}}
\renewcommand{\thefigure}{S\arabic{figure}}

\begin{figure*}\center
	\includegraphics[width=1\linewidth]{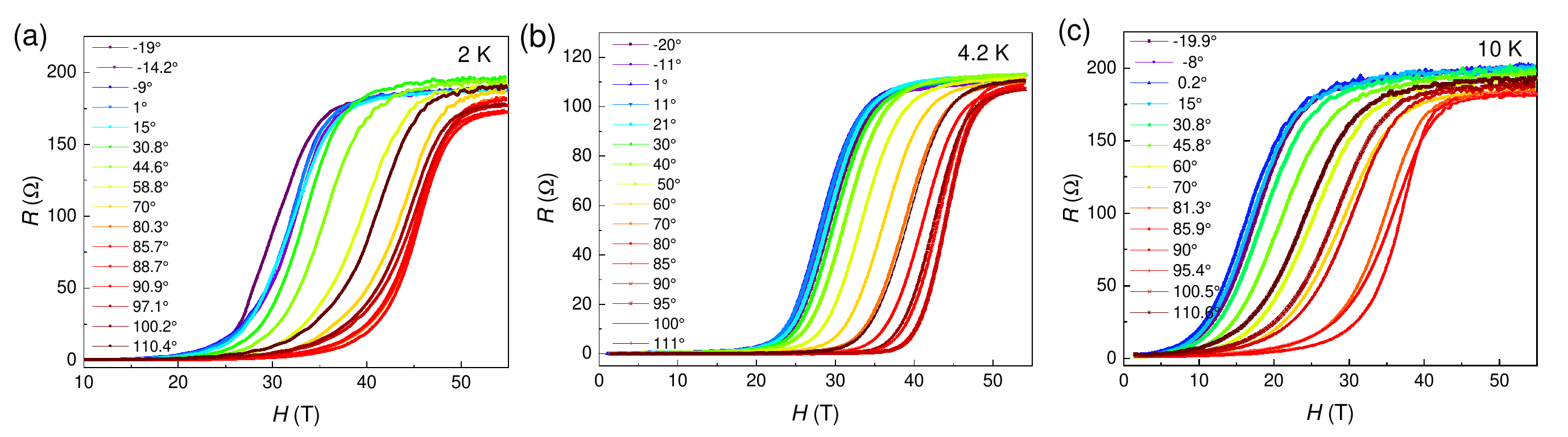}
	\caption{Magnetoresistance at different angles at \textbf{(a)} 2 K, \textbf{(b)} 4.2 K and \textbf{(c)} 10 K respectively.}\label{}
\end{figure*}

\begin{figure*}\center
	\includegraphics[width=0.8\linewidth]{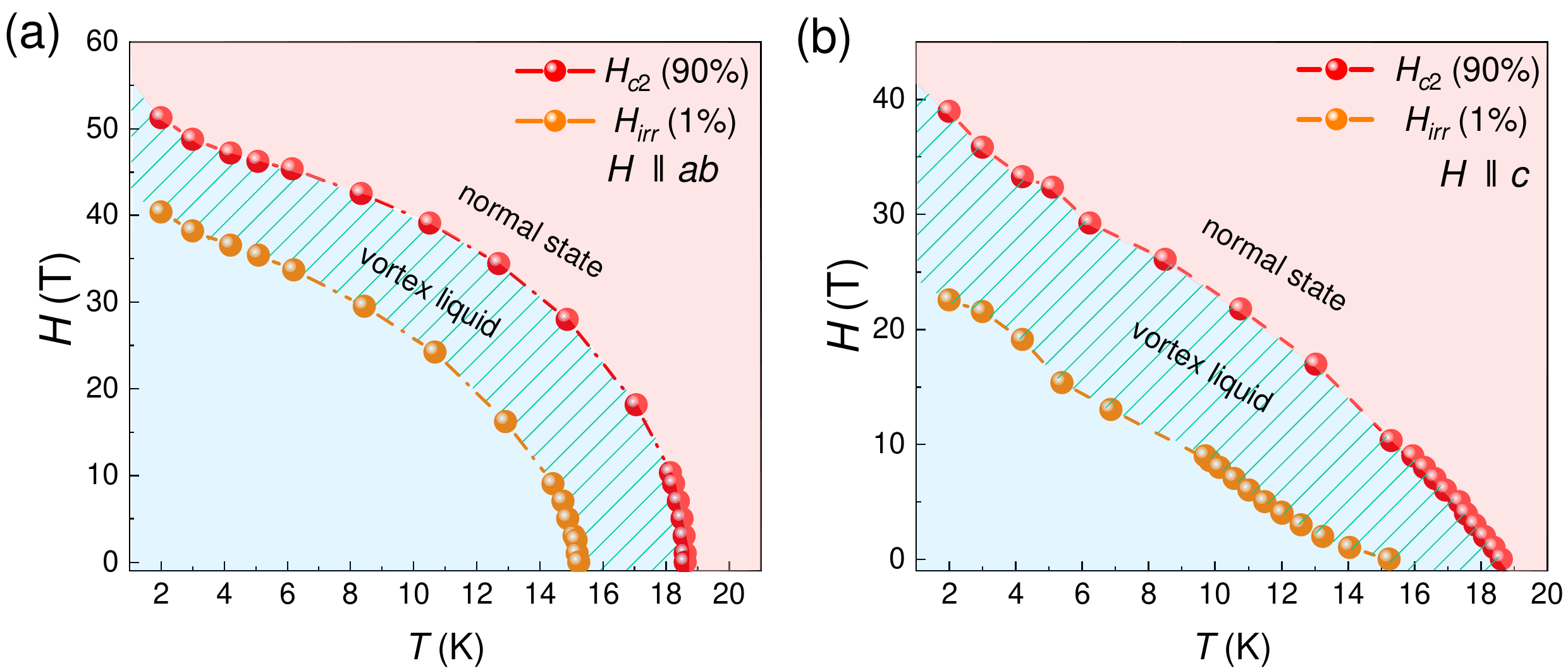}
	\caption{Vortex phase diagram for La$_{0.8}$Sr$_{0.2}$NiO$_{2}$ thin film with (a) $H$ $\Arrowvert$ $ab$ and (b) $H$ $\Arrowvert$ $c$. $H_{\rm{c2}}$ represents the upper critical field obtained from the 90$\%$ of the normal state resistance. $H_{\rm{irr}}$ is an irreversible field obtained by 1$\%$ of the normal resistance.}\label{}
\end{figure*}
The magnetoresistance measurements of La$_{0.8}$Sr$_{0.2}$NiO$_{2}$ at different angles are shown in Figs. S1(a-c), (a) $ T $ = 2 K, (b) $ T $ = 4.2 K and (c) $ T $ = 10 K. The polar angle $ \theta $ is defined as the angle between the field and the $ ab $ plane.

In Fig. S2, we determined $H_{\rm{c2}}$ and irreversibility field $H_{\rm{irr}}$ by taking the criterions of 90$\%$ $R_{\rm{n}}$ and 1$\%$ $R_{\rm{n}}$, and construct the vortex phase diagram. The following three different regimes are clearly distinguishable: (1) Normal state, the region above $H_{\rm{c2}}$. (2) Vortex-liquid phase, which governs the region between the $H_{\rm{irr}}$ line and the $H_{\rm{c2}}$ line. (3) Vortex-glass state. The relatively large $H_{\rm{irr}}$ of our high-quality thin film confirms the ability to carry lossless current under strong magnetic field, which indicates the great application prospect.

\end{document}